# A Brief Review of Data Mining Application Involving Protein Sequence Classification

Suprativ Saha and Rituparna Chaki

Department of Computer Sc. and Engineering
West Bengal University of Technology
Saltlake, Kolkata, India
{reach2suprativ,rituchaki}@gmail.com

**ABSTRACT**

Data mining techniques have been used by researchers for analyzing protein sequences. In protein analysis, especially in protein sequence classification, selection of feature is most important. Popular protein sequence classification techniques involve extraction of specific features from the sequences. Researchers apply some well-known classification techniques like neural networks, Genetic algorithm, Fuzzy ARTMAP, Rough Set Classifier etc for accurate classification. This paper presents a review is with three different classification models such as neural network model, fuzzy ARTMAP model and Rough set classifier model. A new technique for classifying protein sequences have been proposed in the end. The proposed technique tries to reduce the computational overheads encountered by earlier approaches and increase the accuracy of classification.

**Keywords.** Data Mining; Neural Network Model; Fuzzy ARTMAP Model; Rough Set Classifier

## 1 INTRODUCTION

The introduction of new technologies such as computers, satellites and many others has lead to an exponential growth of collected data in many areas. Traditional data analysis techniques often fail to process large amounts of data efficiently. In this case data mining technology can be used to extract knowledge from large amount of data. Recently, the collection of biological data like protein sequences, DNA sequences etc. is increasing at explosive rate due to improvements of existing technologies and the introduction of new ones such as the microarrays. Traditional data analysis techniques are unable to extract meaningful information from the huge amount of biological data sequences, such as the DNA, protein etc. One important area of research is to classify protein sequences into different families, classes or sub classes. One of the emerging techniques for handling such sequences is data mining.

Classification is the most important technique to identify a particular character or a group of them. Different classification methods or algorithms have been proposed by different researchers to classify the protein sequences. The Protein sequence consists of twenty different amino acids which are arranged in some specific sequences. Popular protein sequence classification techniques involve extraction of specific features from the sequences. These features depend on the structural and functional properties of amino acids. These features are compared with their predefined values. Researchers apply some well-known classification techniques like neural networks, Genetic algorithm, Fuzzy ARTMAP, Rough Set Classifier etc. Till date, none have achieved 100% accuracy level. This paper presents a comprehensive study of the on-going research on protein sequence classification followed by a comparative analysis.

The rest of the paper is organized as follows: Section 2 presents a review of classification models; section 3 consists of a comparative analysis, followed by a proposed work in section 4. Section 5 presents the conclusion.

## 2    REVIEW

Different classification techniques have been used to classify protein sequence into its particular class, sub class or family. All these methods aim to extract some features, match the value of these features and finally classify the protein sequence. This paper focuses on mainly three types of classification techniques, the (i) Neural network Model, (ii) the Fuzzy ARTMAP Model, and (iii) the Rough Set Classifier.

### 2.1    NEURAL NETWORK MODEL

Generally there are different types of approaches available for classification, such as decision trees and neural networks. Extracted features of protein sequences are distributed in a high dimensional space with complex characteristics, which is difficult to satisfy model using some parameterized approaches. So neural network based classifier have been chosen to classify protein sequence. Decision tree based techniques fails to classify patterns with continuous features especially as the number of attributes is larger.

Neural network model [1] has been used to classify existing protein sequences. To implement this, some features were extracted from the protein sequences to act as input of this model. A special encoding technique was used to extract high level features from the sequences considering both global and local similarity of the sequences. 2-gram encoding method and 6-letter exchange group methods were used to find global similarity. For local similarity, user defined variables Len, Mut, and occur were used. Minimum description length (MDL) principle was also used to calculate the significance of motif. Some predefined features were used as intermediate layers or hidden layers of the neural network. This model produces 90% to 92% accuracy.

In [2] authors want to classify the protein sequences using neural network model. Here n-gram encoding method (n = 2, 3, 4… N and N = length of the input sequence) was used to extract feature which was applied to construct the pattern matrix. At the end with the help of neural network model this new constructed pattern was matched with the predefine pattern of protein super families or families. N-gram encoding method includes all 2-gram, 3-gram, etc encoding method, so to form the pattern matrix of features extracted from n-gram encoding method, individual also needed. This method involves huge time complexity in case of large sequences. The accuracy level remains 90% only.

[3] Proposes an advancement of the techniques proposed in [2]. At first 2-gram encoding method is applied and using only this result pattern matrix is build. If this matrix is unable to classify the input protein sequence, result of 3-gram encoding method is added to the pattern matrix. The result is then matched using neural network. The performance of this technique is largely dependent on the number of encoding operations performed. In case all the sub patterns are to be checked, performance deteriorates sharply. The average performance is slightly better than [2].

In [4] authors used a probabilistic neural network model to classify the protein sequence into different super family. The paper uses self organized map (SOM) network. The SOM networks can be used to discover relationships within a set of protein sequences by clustering them into different groups. Different types of features like Amino acid distribution, 2-gram amino acid distribution, etc, were extracted from the input protein sequence to construct the pattern matrix. According to the unsupervised learning method of neural network input sequences are placed in the $1^{st}$ layer of neural network, then pattern matrix is presented in the hidden layer ($2^{nd}$ layer) for matching with some predefine values. Different outcome results are summarized in the $3^{rd}$ layer. $4^{th}$ or final layer of the probabilistic neural network model produced the final result of classification. The technique failed to produce impressive results in case of unclassified$_p$ and unclassified$_n$ parameters. The use of SOM network also causes hindrance in interpreting the results.

The main limitations of SOM networks for protein sequence classification are its interpretability of the results, and the model selection. SOM is a straight forward method; there is no chance of back propagation. But to reach a particular goal and increase the accuracy level of the classification back propagation is most important technique. In back propagation based model, there is a chance to move to the previous steps.

The problems faced by the SOM based technique in [4] is overcome by back propagation neural network (BPNN) technique in [5]. Here authors use extreme learning machine to classify protein sequence. This extreme learning machine included the advancement of back propagation technique of neural network model. To evaluate the performance of this machine authors extracted some features like 2-gram encoding method and 6-letter exchange method from the input protein sequence. A pattern matrix was formed using those features and used in the extreme learning machine. Finally accuracy level also is measured.

The use of neural network technique normally neural network is good at handling non-linear data (noise data). The protein sequence being linear, use of neural network does not add up. It has been observed that sequences of 20 different amino acids (Protein sequences) were used as working data in this paper. The data being linear, the use of neural network modelling fails to add any extra benefit. The paper fails to take care of noise in protein sequence even through it uses neural network. The model failed to process the physical relationships which are most important in this purpose. Again regarding the accuracy issue, neural network model provide 90%-92% accuracy. Improvement of this accuracy is mostly needed.

## 2.2 FUZZY ARTMAP MODEL

Generally Fuzzy ARTMAP model, a machine learning method is used to classify the protein sequence. The basic difference between neural network model and fuzzy model is that neural network model do not analysis the data individually, it only provide a knowledge based information. On the other hand Fuzzy model calculates the membership value of every feature using membership functions and implements it in the whole model.

This model [6] was implemented to classify the unknown protein sequence into different predefine protein families or protein sub families. This model produced 93% high accuracy than other models using Structural Classification of Protein (SCOP) 1.69 and ASTRAL database 1.69 as the protein database. A cleaning process was also been conducted on the databases. After that different features were extracted from protein sequences, e.g. physic-chemical properties of the sequences. They calculated the molecular weight (W) and the isoelectric point (pI) of the protein sequences, followed by Amino acid composition of the sequences. The hydropathy composition (C), the hydropathy distribution (D) and the Hydropathy transmission (T) also calculated. After extracting all forty different features an unknown protein sequence was used as the input of the Fuzzy ARTMAP model. Some predefine features which were extracted or generated from known protein sequence also used as the unit of classification rules. This model generated the name of family or sub family of the unknown protein sequence as the output which was taken as the input of the Fuzzy ARTMAP model.

In [7] author wants to classify protein sequence using Fuzzy model. Calculating the membership value using the membership function is most important in fuzzy model. At first feature is extracted using 6-letter exchange group method. Then membership value is assigned and constructs the pattern matrix. Using a fuzzy rule pattern matrix was distributed into 3 small groups (i) small, (ii) medium and (iii) large. Now according to the target, choose a group and further distribution was done to reach to goal. At the end, the model is tested using uniport 11.0 dataset which contain globin, kinase, ras and trypsin super families of protein. In this paper number of antecedent variables is huge. It is right that increase of antecedent variable, increase the classification accuracy but it also increases the CPU time.

[8] Proposes an advancement of the techniques proposed in [7]. This technique tries to decrease the CPU time without changing the classification accuracy. Here features also extracted using 2-gram encoding method and 6-letter exchange group method and according to the membership value of features pattern matrix was distributed into 3 small groups. But executing the distribution method a new algorithm is applied on the value of features. This algorithm provides a rank on the value of the features using the feature ranking algorithm and according to the rank features is arranged in descending order. Now collect the top ranked features to construct the pattern matrix. In this way this technique can reduce the CPU overhead. This is a normal, easy, human understandable and alignment – independent method. As a result every biologist can easily understand this method and feel free to implement it. At the end this method is evaluated and compared to the non fuzzy technique (C 4.5). The computational complexity is reduced, but the accuracy level remains the same as the earlier method [7].

Fuzzy modeling helps in the data analysis although storage and time requirement are high. The construction of fuzzy sets, for every iteration adds up to the computational complexity as well. This model also failed to process the physical relationships which are most important in this purpose.

## 2.3 ROUGH SET CLASSIFIER

Generally machine learning methods such as the neural network model, Fuzzy ARTMAP model etc., are insufficient to handle large number of unnecessary features, extracted for rule discovery [11]. As a result they try to select the features to reduce the computation time. But these methods also degrade their performance. Accuracy level is not sufficient since every feature is equally important for proper classification. Rough set classifier is a new model to overcome this problem.

Rough sets theory is a machine learning method, which is introduced by Pawlak [10] in the early 1980s. It implements the concept of set theory to make some decision. The indiscernibility relation that induced minimal decision rules from training examples is the important notation in rough set model. To identify the minimal set of the features, if-else rule is used on the decision table.

This new classification model [9] can classify the voluminous protein data based on structural and functional properties of protein. This model is faster, accurate and efficient classification tool than the others. Rough Set Protein Classifier provides 97.7% accuracy. It is a hybridized tool comprising Sequence Arithmetic, Rough Set Theory and Concept Lattice. It reduces the domain search space to 9% without losing the potentiality of classification of proteins. An innovative technique viz., Sequence Arithmetic (SA) to identify family information and utilize it for reducing the domain search space is proposed. Rules are generated and stored in Sequence Arithmetic database. A new approach to compute predominant attributes (approximate reducts) and use them to construct decision tree called Reduce based Decision Tree (RDT) is proposed. Decision rules generated from the RDT are stored in RDT Rules Database (RDTRD). These rules are used to obtain class information. The infirmity of RDT is overcome by extracting spatial information by means of Neighbourhood Analysis (NA). Spatial information is converted into binary information using threshold. It is utilized for the construction of Concept Lattice (CL). The Associated Rules from the CL are stored in Concept lattice Association Rule Database (CARD). Further, the domain search space is confined to a set of sequences within a class by using these Association Rules. Time complexity of this model is $O(n) + O(f) + O(\log C) + O(2^r)$, where the unknown sequence y with size n. No of the families in the database is f. 'C' is the number of classes in a given family and 'r' is the number of proteins in classes then the CL will have $2^r$ nodes.

In [11] authors use rough set classifier to extract all the features necessary for classification. The feature set was built based on compositional percentages of the 20 amino acids properties. The authors had used Rosetta system for data mining and knowledge discovery. In the first phase, a method is implemented on the whole datasets in which all the subfamilies were included ignoring the small size of sequences. Rough set model generally use standard Genetic Algorithms. The Rough Set was further applied to classify the data and evaluate the performance of the seven subfamilies. This paper achieves a satisfactory accuracy level without increase the computational time.

The Rough Set Classifier model provides knowledge based information only without any analysis of data. For properly classifying protein sequences, both play an important role. Instead of classifying protein sequence into classes or sub classes, this model provides a small known sequence from a long unknown protein sequence. Thus it requires extra time and space for further classification of the output sequence into classes or sub classes. The accuracy level is 97.7%, which leaves scope for improvement.

## 3 COMPARATIVE ANALYSIS

| Techniques | Neural Network based Classifier [1,2,3,4,5] | Fuzzy ARTMAP based Classifier [6,7,8] | Rough Set based Classifier [9,10,11] |
|---|---|---|---|
| Database Uses | The Int. Protein Seq. Database Release 62 | i) SCOP 1.69 <br> ii) ASTRAL 1.69 | NCBI (Blast) |
| Features Selection | *1) Global similarity* <br> i) 2-gram encoding method <br> ii) 6-letter exchange group methods. <br> *2) Local similarity* <br> i) Len, Mut, and occur calculation. <br> ii) Min. description length (MDL) principle | i) Molecular weight (W) <br> ii) Isoelectric point (pI) <br> iii) Hydropathy composition (C) <br> iv) Hydropathy distribution (D) <br> v) Hydropathy transmission (T) | i) Sequence Arithmetic <br> ii) Reduce based Decision Tree (RDT) <br> iii) Neighbourhood Analysis (NA) <br> iv) Concept Lattice (CL) |
| Accuracy Level | 90% to 92% | 93% | 97.7% |
| Drawbacks | i) Better for Non-linear and Noisy data. <br> ii) Does not handle Physical relationship. | i) Concerned only about the physical Structure of AA. <br> iii) Does not handle Physical relationship. | i) No analytical output. <br> ii) Need Extra Time and Space |

## 4  PROPOSED MODEL

The main purpose of this model is to classify the unknown protein sequence in to different families, classes or sub classes with high accuracy level and low computational time. To implement this goal choose a unknown protein sequence as an input and extracts some features from it and match with predefine values to classify the sequence into classes, sub classes and families. To do this first of all, extraction of features which is used to classify, is very important.

The proposed technique consists of three phases. The first phase aims to reduce the input dataset. The second phase helps to increase accuracy level of classification and the third phase implies the association rule to classify the protein sequence. Figure 1 gives a pictorial representation of the different modules of the proposed technique.

### 4.1  PHASE 1

2-gram encoding method and 6-letter exchange group method both are used to extract the global similarity of the protein sequence [1]. These two methods are directly related to the structure of a protein sequence. So, to extract the knowledge based information, we have to calculate the global similarity of protein sequence. In the first phase if we extract the knowledge based information using the back propagation technique of the neural network [5] then we are able to reduce the total dataset. So those two methods are performed at first to construct the pattern matrix. Now this pattern matrix acts as the input of the neural network model. Here we use only two techniques to extract the feature, so number of the features is very less. In this situation it is right that we do not reach the require accuracy but we can be able to reduce the total dataset for further classification within low computational time.

### 4.2  PHASE 2

In the second phase, after reduce the dataset, Molecular weight, isoelectric point, Hydropathy composition, Hydropathy distribution, Hydropathy transmission will be calculated to extract the features.. After extracting those features, a feature ranking algorithm will be applied on it. This algorithm will be able to provide a rank to the feature values and arrange the features values according to the descending order [8]. Top rank means have an extra ability to classify the protein sequence. After that pattern matrix will be generated using top ranked in the second highest level of reduce based decision tree. Now this pattern matrix will be distributed within three groups [7] and applied to the Fuzzy model. Those features which are used here generally deal with the molecular structure of the amino acids. So it is possible to eliminate huge no of classes, sub classes and families of protein in which the input protein sequence does not belongs. In this case data by data analysis was implemented instead of extracting knowledge based information. If our dataset is huge then data by data analysis takes huge computational time. But here our data set is small because in the first stage we are able to reduce the data set. The main advantage of the data by data analysis is it provides the high accuracy level.

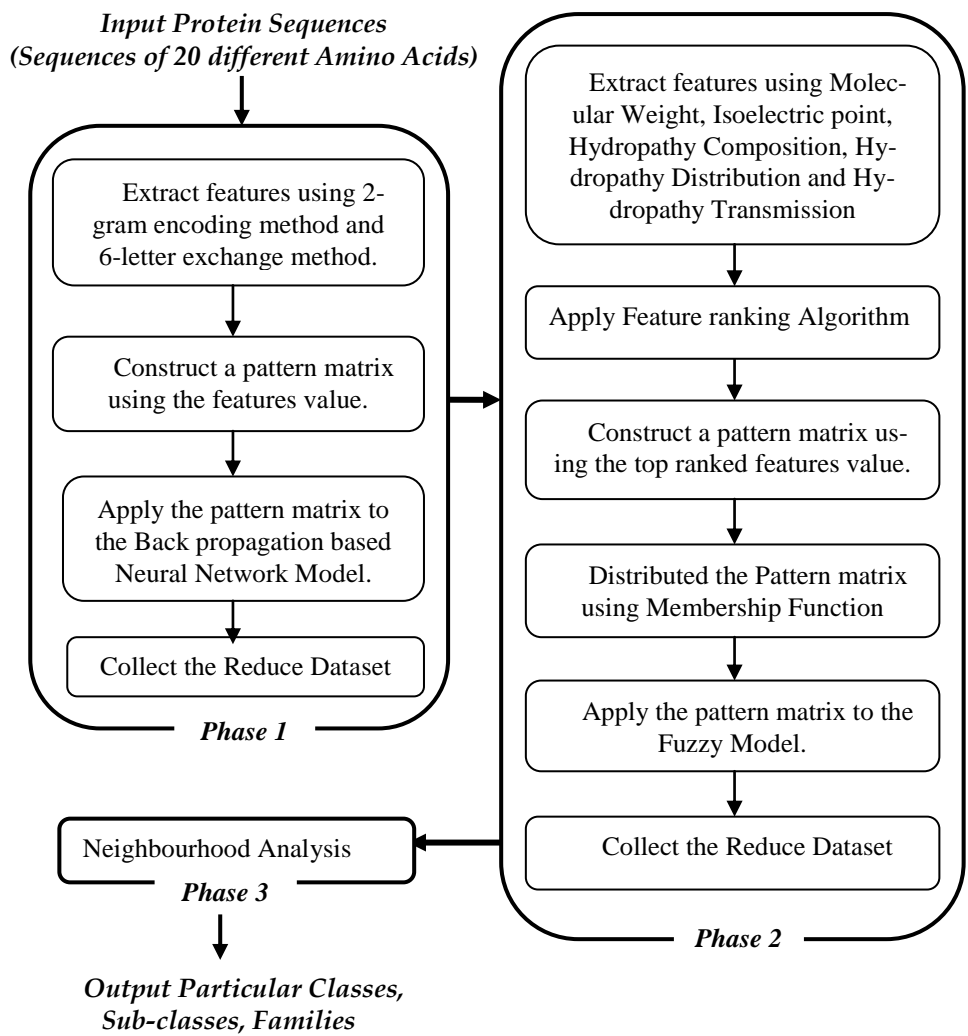

**Fig. 1.** Pictorial representation of the different modules of the proposed technique

### 4.3 PHASE 3

In the third and final phase, Neighbourhood Analysis (NA) will be used to classify the input sequence in the particular class or family. To use neighbourhood analysis we generally apply association rule. This rule has a power to extract the particular association between the protein sequence and classes, sub classes and families. So it is possible to eliminate all other classes, sub classes and families of protein in which this input protein sequence do not belongs

## 5   CONCLUSION

In the recent treads, analysis the large amount of biological data like protein sequences is very difficult using traditional database system. In this case data mining technique can be used to classify the unknown protein sequence. But the different models, which are used to classify the protein sequence is not perfect regarding the both accuracy level and computational time. This dissertation includes a detail review of ongoing research work involving three different techniques to classify the protein sequences. It has been observed that knowledge based and analysis of data form integral parts of protein sequence classification. The accuracy level of each proposed model has been studied. Finally, a new classification model was proposed which can classify the unknown protein sequences into families, classes or sub classes, producing knowledge based information beside data analysis technique. In future different analysis will be done with this new proposed classification model.